\documentclass[runningheads]{llncs}
\usepackage[T1]{fontenc}
\usepackage{graphicx}
\usepackage{enumitem}
\usepackage{amssymb}
\usepackage{amsmath}
\usepackage{booktabs}
\usepackage[colorlinks=true,
            linkcolor=blue,     
            citecolor=blue,     
            urlcolor=black,      
            anchorcolor=blue]{hyperref}
\usepackage[misc]{ifsym}
\usepackage{cite}
\begin{document}
\author{
Kai Cheng\inst{1} \and
Hao Wang\inst{1}\textsuperscript{(\Letter)} \and
Wei Guo\inst{2} \and
Weiwen Liu\inst{3} \and
Yong Liu\inst{2} \and
Yawen Li\inst{4} \and
Enhong Chen\inst{1}\textsuperscript{(\Letter)}
}
\institute{
University of Science and Technology of China, Hefei, China\\
\email{\{ck2020\}@mail.ustc.edu.cn}\\
\email{\{wanghao3,cheneh\}@ustc.edu.cn}
\and
Huawei, Shanghai, China\\
\email{\{guowei67,liu.yong6\}@huawei.com}
\and
Shanghai Jiao Tong University, Shanghai, China\\
\email{wwliu@sjtu.edu.cn}
\and
Beijing University of Posts and Telecommunications, Beijing, China\\
\email{warmly0716@126.com}
}

\title{Efficient Personalized Reranking with Semi-Autoregressive Generation and Online Knowledge Distillation}

\titlerunning{Efficient Personalized Reranking via SAG and OKD}
\authorrunning{K. Cheng et al.}
%
%

%
\maketitle              

\begin{abstract}
Generative models offer a promising paradigm for the final-stage reranking in multi-stage recommender systems, with the ability to capture inter-item dependencies within reranked lists. However, their practical deployment still faces two key challenges: (1) an inherent conflict between achieving high generation quality and ensuring low-latency inference, making it difficult to balance the two, and (2) insufficient interaction between user and item features in existing methods.
To address these challenges, we propose a novel \textbf{P}ersonalized \textbf{S}emi-\textbf{A}utoregressive with online knowledge \textbf{D}istillation (PSAD) framework for reranking. In this framework, the teacher model adopts a semi-autoregressive generator to balance generation quality and efficiency, while its ranking knowledge is distilled online into a lightweight scoring network during joint training, enabling real-time and efficient inference. Furthermore, we propose
a User Profile Network (UPN) that injects user intent and models interest dynamics, enabling deeper interactions between users and items.
Extensive experiments conducted on three large-scale public datasets demonstrate that PSAD significantly outperforms state-of-the-art baselines in both ranking performance and inference efficiency.

\end{abstract}

\section{Introduction}
Large-scale online platforms widely adopt Multi-Stage Recommender Systems (MRS) to enhance user experience and optimize decision-making~\cite{onerec,new-gen3,wang2025mf,wang2019mcne,wang2021hypersorec,wang2025generative,wang2025universal,wang2025dlf,wang2025enhancing,wang2024denoising}. Such systems typically consist of three stages: retrieval, ranking, and reranking~\cite{relife,zhang2024unified,zhang2025killing,zhang2025td3,zhang2025rag,zhou2026survey,zhou2025multi,zhang2026next,zhang2026can,zhang2026thinking}, where the reranking stage refines a small set of candidate items for final presentation~\cite{hron2021component,wu2024survey}. Unlike the previous stages, reranking must account for both the individual relevance of items and their mutual dependencies to produce an optimal overall list~\cite{liu2022neural}. Effectively modeling interactions between items and understanding diverse user interests remain key challenges in this field~\cite{feng2021revisit}.

Several research efforts have been dedicated to reranking. Early studies adopted discriminative approaches, which rank items based on their individual scores, primarily model relationships within the candidate set~\cite{MIDNN,li2022pear}. Consequently, they often neglect the crucial contextual dependencies within the final reranked list, leading to suboptimal performance~\cite{seq2slate,new-gen3}.

\begin{figure}
	\centering
	\setlength{\belowcaptionskip}{-0.0cm}
	\setlength{\abovecaptionskip}{-0.0cm}
    \vspace{0mm}
         \includegraphics[width=\textwidth]
    {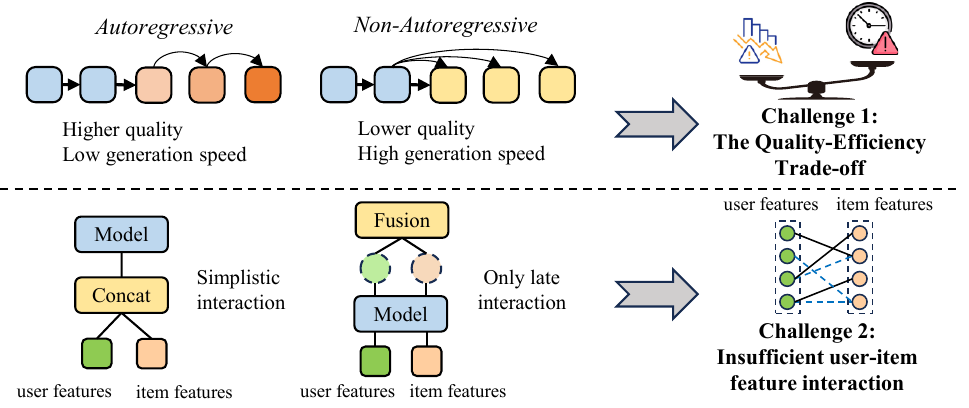}
    \vspace{0mm}
	\caption{Two challenges faced by generative reranking methods. }
	\label{fig:ag}
	\vspace{-1mm}
\end{figure}

In recent years, generative models have gained significant attention for their ability to dynamically generate item permutations while explicitly modeling dependencies within the reranked list~\cite{huzhang2021aliexpress,ai2019learning}, ultimately optimizing the final display list.
This approach includes both autoregressive and non-autoregressive strategies~\cite{MIR}: the former generates items sequentially based on previously generated ones~\cite{seq2slate,gong2022real}, while the latter produces the entire sequence without strict order, focusing on global item relationships~\cite{nar4rec,lin2024discrete}.

Despite notable progress, generative reranking methods still encounter several key challenges, are illustrated in Fig.~\ref{fig:ag}:
(1) \textbf{The Quality-Efficiency Trade-off}: Autoregressive models achieve high generation quality through fine-grained sequential modeling but suffer from slow inference and error accumulation due to their step-by-step generation process~\cite{nar4rec}. In contrast, non-autoregressive models offer efficient parallel generation but rely on a strong conditional independence assumption, resulting in incoherent outputs~\cite{new-gen4}. To mitigate these quality issues, researchers often employ iterative refinement~\cite{new-gen3} or two-stage paradigm~\cite{nar4rec}, which in turn compromise their original efficiency and real-time capability.
(2) \textbf{Insufficient User–Item Feature Interaction}: Although existing methods have explored personalized modeling~\cite{relife,PIER}, the interaction between user and item features remains insufficient. Some approaches simply concatenate user and item features~\cite{PRM}, failing to capture the semantic variations of the same item under different user perspectives; others perform interaction only after extracting high-level representations~\cite{MIR}, overlooking early-stage latent connections and thus limiting the modeling of complex user interest patterns.

To address these issues, we propose a novel \textbf{P}ersonalized \textbf{S}emi-\textbf{A}utoregressive with online knowledge \textbf{D}istillation (\textbf{PSAD}) framework for reranking.
First, to tackle Challenge (1), we introduce a unified online knowledge distillation framework. We employ a semi-autoregressive generation paradigm whose block-wise mechanism reduces error accumulation while significantly lowering training overhead. To further reduce inference latency, we perform online knowledge distillation, jointly training a lightweight scoring network and the generator from scratch with shared encoder parameters. Unlike offline methods that require a pretrained teacher, our approach allows the scoring network to learn the generator's ranking knowledge on-the-fly without incurring significant distillation overhead, thereby approximating high performance with low latency at inference. 
Second, to tackle Challenge (2), we deeply inject user preferences into the model via our User Profile Network (UPN). It comprises two components: personalized gates that dynamically adapt the semantic representations of items based on user profiles, and a personalized position encoding mechanism to capture unique user interest decay patterns.
In summary, our contributions are as follows:
\begin{itemize}[leftmargin=*,align=left]
\item We developed PSAD, which is the first to effectively address latency and quality issues in generative ranking from a novel perspective of fine-grained user-item feature interactions in low-latency generative reranking.
\item We design an innovative online distillation architecture: a semi-autoregressive teacher achieves high quality and training efficiency via block-wise generation, while distilling its knowledge on-the-fly to a lightweight student for significantly reduced inference latency.
\item We design UPN, consisting of personalized gates and a personalized position encoding mechanism, to achieve deep fusion of user and item features.
\item Extensive experiments demonstrate that PSAD significantly reduces inference latency while maintaining high-quality ranking performance.
\end{itemize}

\section{Related Work}
\subsection{Reranking Method}
Reranking methods refine an initial list of candidates to optimize overall ranking quality. Early discriminative approaches, such as the RNN-based DLCM~\cite{DLCM}, the Transformer-based PRM~\cite{PRM} and SetRank~\cite{pang2020setrank}, and the GNN-based IRGPR~\cite{liu2020personalized}, primarily focused on modeling contextual dependencies within the candidate set. In recent years, generative reranking has emerged as a new paradigm, framing the task as a process of "generating the optimal permutation." Representative methods like Seq2Slate~\cite{seq2slate}, CRUM~\cite{crum}, and GRN~\cite{feng2021grn} employ autoregressive or reinforcement learning strategies to directly generate the optimized item sequence; PIER~\cite{PIER} leverages hashing algorithms to filter candidates, while others like NAR4Rec~\cite{nar4rec} introduce non-autoregressive generation to reduce inference latency.
However, this emerging paradigm also faces new challenges. Existing methods generally struggle to balance generation quality with inference efficiency: while autoregressive models can produce high-quality results, their inference is slow; conversely, non-autoregressive models are highly efficient but often lack coherence due to overly strong independence assumptions, making it difficult to achieve ideal performance in practical deployments.

\subsection{Personalized Reranking Method}
Mining user intentions from behavioral data has become a key direction for personalized reranking. Existing methods, such as PEAR~\cite{li2022pear} and MIR~\cite{MIR}, model historical lists to capture user preferences but often overlook the full utilization of user features. Some models, like GRN~\cite{feng2021grn} and PRM~\cite{PRM}, simply concatenate user and item features, while others, such as MIR~\cite{MIR} and RAISE~\cite{lin2022attention}, perform user–item interactions only in later hidden layers. This delayed interaction limits the model’s ability to capture complex user–item relationships in a timely manner, thereby constraining the effectiveness of personalized modeling.

\subsection{Semi-autoregressive Generation}
To improve the efficiency of autoregressive models, early NLP research introduced semi-autoregressive modeling, which generates multiple token spans per step instead of a single token~\cite{semi1}. Later advancements, such as parallel decoding~\cite{semi2} and speculative decoding~\cite{semi3}, further refined this idea.
Inspired by these developments, we apply the semi-autoregressive generation mechanism to the reranking stage of recommender systems, enabling parallel generation of multiple candidate items to enhance overall performance.

\section{Preliminary}
\vspace{-5pt}
\paragraph{\textbf{Definition1. (Reranking Problem)}}
The reranking stage is typically defined as follows: given a user \( u \), the candidate sequence \( C_u \) obtained from the previous recommender stage, the user's recent interaction history \( H_u \), and the user's profile features \( \mathcal{P}_u \) (e.g., age, gender, etc.). Each item in the \( C_u \) or \( H_u \) set is characterized by sparse features (such as category ID, brand ID, etc.) and dense features (such as price, score, etc.). The reranking problem involves learning a mapping function $ \mathcal{F} $, such that \( \mathcal{F}(C_u, H_u, \mathcal{P}_u) \to R_u \), where \( R_u \) is the reranked list presented to the user and \( R_u \subseteq C_u \).  The length of \( H_u\) is denoted as \(N\), the \( C_u \) as \(M\), and \( R_u \) as \(T\), which typically satisfies \(T \leq M\).

\vspace{-5pt}
\paragraph{\textbf{Definition 2. (Generative Reranking)}} 
Generative reranking models aim to find the optimal sequence $R_u$ by directly modeling its conditional probability. 
The goal is to generate the list that maximizes the likelihood given the user's context:
$
R_u^* = \arg\max_{R_u \in \mathcal{P}(C_u)} p(R_u | C_u, H_u, \mathcal{P}_u),
$
where $\mathcal{P}(C_u)$ represents the space of all valid permutations of items from the candidate set $C_u$. 
Different modeling assumptions on $p(R_u \mid \cdot)$ lead to various generation paradigms, such as autoregressive or non-autoregressive approaches.

\section{Methodology}
\begin{figure*}[t]
	\centering
	\setlength{\belowcaptionskip}{-0.0cm}
	\setlength{\abovecaptionskip}{-0.0cm}
	\includegraphics[width=0.99\textwidth]{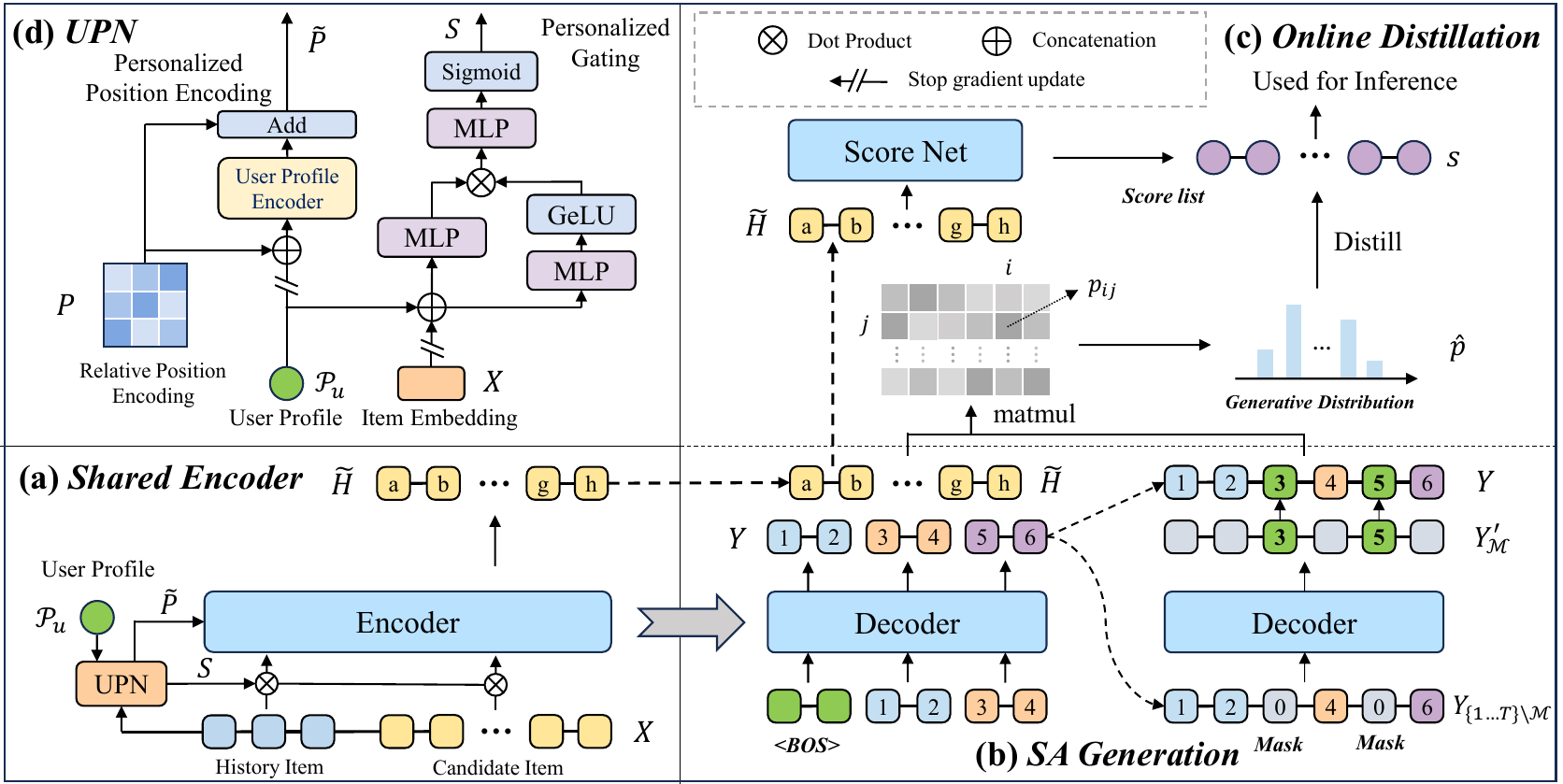}
    \vspace{2mm}	
    \caption{The overall framework of PSAD and its sub-modules.}
	\label{fig: framework}
	\vspace{-3mm}
\end{figure*}
\vspace{-5pt}
Existing generative reranking models often struggle to balance generation quality and inference efficiency while lacking sufficient depth in personalized modeling. To address these challenges, we propose PSAD, whose overall architecture is illustrated in Fig.~\ref{fig: framework}. This section provides a detailed description of its core components: 
First, a shared encoder (Fig.~\ref{fig: framework}(a)) processes all input sequences, as described in Sec. \ref{encoder}. Next, we address the trade-off between quality and efficiency through semi-autoregressive generation (Fig.~\ref{fig: framework}(b)) and online distillation (Fig.~\ref{fig: framework}(c)), detailed in Sec. \ref{generator}. Then, the User Profile Network (UPN) (Fig.~\ref{fig: framework}(d)) is introduced to enable deep personalization, as discussed in Sec. \ref{usernet}. Finally, Sec. \ref{model_inference} presents the inference process of PSAD.

\subsection{Encoder} \label{encoder}

\paragraph{\textbf{Embedding Layer}}
According to \textit{Definition~1}, for each item $v$ in the user's historical sequence $H_u$ or the candidate set $C_u$, 
the $i$-th sparse feature of the item, is mapped into representation $e_{v,i}$ through the corresponding embedding table. 
Subsequently, we concatenate the embeddings of all sparse features with the item's dense features to form the final item embedding:
$x_v = [e_{v,1} \oplus \dots \oplus e_{v,k} \oplus x_d^v] \in \mathbb{R}^{d_x},$
where $k$ denotes the number of sparse features and $x_d^v$ represents the dense feature vector. 
The operator $\oplus$ indicates vector concatenation. 
By stacking all item embeddings $x_v$, we obtain the input item matrix 
$ X \in \mathbb{R}^{(N+M) \times d_x}.$
Similarly, the embedding vector of user $u$ is represented as 
$ \mathcal{P}_u \in \mathbb{R}^{d_u}.$

\paragraph{\textbf{Shared Encoder}}
To better capture the relationships among items, we adopt a self-attention structure, and the attention score matrix is calculated by:
\begin{equation} \label{eq1}
A^l = \frac{\text{Softmax}((X^{l} W^l_Q)(X^{l} W^l_K))}{\sqrt{d_x}}+P^l,
\end{equation}
where \( l \) denotes the $l$-th layer in the model, \( W_Q \) represents the learnable parameters \( \in \mathbb{R}^{d_x \times d_x} \), and similarly for the others.
The matrix \( P^l \) is a learnable parameter, where \( P^l_{i,j} \) represents the positional encoding value for the \( i \)-th and \( j \)-th positions.
The attention score is then multiplied by the attention value and passed through a Feed-Forward layer to obtain the output of the encoder:
\begin{equation}
H^l = A^l (X^l W_V^l), \quad \tilde{H}^l = \text{FFN}(H^l).
\end{equation}
The encoder stacks multiple identical layers, where the input $X^l$ of each layer is the output $\tilde{H}^{l-1}$ from the previous layer.

\subsection{Generator and Online Distillation}
\paragraph{\textbf{Semi-Autoregressive Generation}} \label{generator}


After the encoding stage, we employ a generator to rerank the candidate set. To overcome the inherent limitations of autoregressive and non-autoregressive models, we design a semi-autoregressive generation scheme that reduces the number of generation steps to mitigate error accumulation and improve generation efficiency, while retaining autoregressive dependencies to capture sequential relationships. Conceptually, it is analogous to a pointer network~\cite{seq2slate}, where query vectors are generated to select items from the candidate set. In practice, the generator constructs blocks containing $K$ items in parallel at each iteration step until the complete sequence is formed.

The specific process is as follows: let the target result length be $T$, and the block size be $K$, the position indices for the $b$-th block are defined as
$\mathcal{I}_b = \{ (b-1)K + 1, \dots, \min(bK, T) \}.$
For each block, we generate \( K \) tokens simultaneously, conditioned on the previously generated context:
\begin{equation}
Y^{l}_{\mathcal{I}_b} = \text{CrossAttn}( H^l, Y_{<\mathcal{I}_b}^{l}),
\end{equation}
where \( \text{CrossAttn} \) denotes the model cross attention, and \( Y_{<\mathcal{I}_b} \) represents the result from tokens before the current block.
To reduce inconsistencies inside each parallelly generated block, we employ a contextual enhancement step that refines the full sequence $Y$ using a mask-and-refine paradigm. We randomly mask a subset of tokens $\mathcal{M}$, where $|\mathcal{M}| = \gamma$, and then update them with predictions conditioned on the unmasked context:
\begin{equation}
Y'_{\mathcal{M}} = \text{CrossAttn}( H^l, Y_{\{1..T\} \setminus \mathcal{M}} ), \quad Y \leftarrow Y_{\{1..T\} \setminus \mathcal{M}} \cup Y'_{\mathcal{M}}.
\end{equation}

The generation probability matrix is derived from the multiplication of the generated sequence representations and the hidden vectors of the candidate set:
\begin{equation} \label{infergen}
p_{ij} = \text{Sigmoid}(\tilde{H}^L_i \cdot Y^L_j),
\end{equation}
where \( L \) represents the total number of layers. The matrix element \( p_{ij}\) represents the probability that the \( i \)-th item in the candidate set is the \( j \)-th position in the reranked list. Then we compute the generative loss:
\begin{equation}\small
\mathcal{L}_{\textit{gen}} = \sum_j \omega_j (\sum_i (-\hat{y}_{i} \log p_{ij})+\max\{0, 1-\min_{i:\hat{y}_{i}=1}p_{ij}+\max_{i:\hat{y}_{i}=0}p_{ij}\}),
\end{equation}
where $\hat{y}_{i}$ is ground truth where $\hat{y}_{i}$ is 1 if $ i $-th item with positive label otherwise 0.
Our generative loss combines a cross-entropy term for pointwise accuracy and a hinge loss for listwise ranking order, weighted by a DCG inspired term, $\omega_j = \frac{1}{\log(j + 1)}$, to prioritize early positions.


\paragraph{\textbf{Online Knowledge Distillation}}
To further reduce the computational burden during inference, we propose a unified Online Knowledge Distillation framework. This framework includes a lightweight Scoring Network as the student model, which is jointly trained with the previously introduced generator serving as the teacher model. The scoring network computes a score $s_i$ for each candidate item $i$ and is supervised using a standard cross-entropy loss:
\begin{equation} \label{inferdis}
s_i = \text{Sigmoid}(\text{MLP}(\tilde{H}^L_i)), \quad \mathcal{L}_{\textit{scorer}} = - \sum_i \hat{y}{i} \log s_{i}.
\end{equation}
Unlike offline distillation methods that rely on a pre-trained teacher model, we perform on-the-fly distillation to transfer the generator’s ranking knowledge to the student model, aggregating the generator’s probability matrix into target scores $\hat{p}_i$ for each item using an exponential decay weighting scheme:
\begin{equation}
\tilde{p}{ij}=\frac{\exp{(p{ij})}}{\sum_k \exp{(p_{kj})} }, \quad \hat{p}{i} = \frac{\sum_j \tilde{p{ij}} \cdot e^{-j}}{\sum_j e^{-j}},
\end{equation}
Subsequently, to more effectively measure the consistency between the teacher's and student's output distributions, we employ Kullback-Leibler Divergence as the distillation loss:
\begin{equation}
\mathcal{L}_{\textit{Distillation}} = \text{KL}\Big(\text{Softmax}(\hat{p}/\tau) \parallel \text{Softmax}(s/\tau)\Big),
\end{equation}
where $\tau$ is the temperature that softens the probability distributions.
Finally, the total training loss consists of the losses from the collaborative training process:
\begin{equation}
\mathcal{L} = \mathcal{L}_{\textit{gen}} + \mathcal{L}_{\textit{scorer}} + \alpha \cdot \mathcal{L}_{\textit{Distillation }},
\end{equation}
where $\alpha$ is a hyperparameter balancing the distillation loss. By optimizing this objective, enabling the student model to emulate the teacher's complex ranking preferences with high efficiency during inference, thus achieving high generation quality while maintaining low-latency inference. 

\subsection{User Profile Network} \label{usernet}
\paragraph{\textbf{Personalized Gating}}
Previous reranking methods typically encode items as fixed embeddings. However, due to the diversity of user interests, item representations should adapt dynamically to user-specific characteristics. To this end, we design a personalized gating unit that integrates user profile information into item representations while preserving the original item features, thereby enabling personalized item embeddings. Its working principle is as follows:
\begin{equation}
\mathcal{G}_{1} = \text{MLP}_1(\oslash X, \mathcal{P}_u), \quad \mathcal{G}_2 = \text{GeLU}(\text{MLP}_2(\oslash X, \mathcal{P}_u)),
\end{equation}
where MLP refers to a fully connected layer, and GeLU is a nonlinear activation function. \( \oslash \) denotes the stop gradient operation, this ensures that the module optimizes only the user profile, thereby facilitating faster convergence during training. 
Subsequently, we compute the interaction between the activated and non-activated representations of the item-user pair to generate the gating signal:
\begin{equation}
S = \text{Sigmoid}(\text{MLP}(\mathcal{G}_1 \odot \mathcal{G}_2)), \quad
\hat{X} = S \odot X,
\end{equation}
where $ \odot $ represents the Hadamard product. The gating signal $S$ is applied to the item embedding $X$, and combined with the projected user profile to produce the personalized representation $\hat{X}$, which serves as the input to Equation \eqref{eq1}.

\paragraph{\textbf{Personalized Position Encoding}}
In Equation \eqref{eq1}, the relative position encoding $P^l$ assumes a fixed interest decay pattern for all users. To enable personalization, we propose a position-adaptive scheme that allows the model to dynamically adjust the universal position bias $P^l_{ij}$ based on the user profile. Specifically, we generate an adaptive bias for each position pair by fusing the user preference $\mathcal{P}_u$ with the generic positional relationship $P^l_{ij}$. The bias is added to the original encoding to obtain the personalized positional encoding $\hat{P}^l$:
\begin{equation} \label{eq3}
\hat{P}^l_{ij} = P^l_{ij} + \text{Tanh}(\text{MLP}(\oslash \mathcal{P}_u \oplus P^l_{ij})).
\end{equation}
This enables the model to learn user-specific interest dynamics rather than a uniform temporal decay pattern.

\subsection{Model Inference} \label{model_inference}
\vspace{-2pt}
 During inference, the generator computes the probability distribution according to Equation \eqref{infergen} and produces candidate sequences via beam search, which are then scored by PRM to select the highest-scoring sequence. The scoring network, in contrast, directly assigns scores based on Equation \eqref{inferdis} and ranks items to select the top $T$ candidates. Evidently, inference via the scoring network is more efficient, playing a crucial role in alleviating latency in the reranking process.

\section{EXPERIMENTS}
\begin{table}[t]
    \caption{Dataset statistics.}
    \centering
    \vspace{-0.2cm}
    \setlength{\tabcolsep}{1mm}
    \begin{tabular}{c|c|c|c}
        \toprule
        \textbf{Dataset} & \textbf{\#Users} & \textbf{\#Items} & \textbf{\#Records} \\
        \midrule
        Ad & $1.06\times10^{6}$ & $8.27\times10^{5}$ & $2.04\times10^{6}$ \\
        PRM Public & $7.44\times10^{5}$ & $7.25\times10^{6}$ & $1.44\times10^{7}$ \\
        Avito & $1.32\times10^{6}$ & $ 2.36 \times 10^{7}$ & $5.36\times10^{7}$\\
        \bottomrule
    \end{tabular}
    \label{tab:dataset}
\end{table}
\subsection{Experimental Setup}
\subsubsection{Datasets}
To ensure the broad applicability of the proposed method, we conducted extensive experiments on public datasets, including the E-commerce Reranking dataset and Ad dataset.
\begin{itemize}[leftmargin=*,align=left]
\item \textbf{Ad\footnote{\url{https://tianchi.aliyun.com/dataset/56}}}: The dataset records interactions between users and advertisements, containing 9 user features (e.g., ID, age, and occupation) and 6 item features (e.g., ID, campaign, and brand). Based on the timestamps of user interactions with advertisements, we transform each user's records into ranking lists. Items interacted with within five minutes are grouped into the same list.
    \item \textbf{ PRM Public\footnote{\url{https://github.com/rank2rec/rerank}}}: A dataset includes 3 user profile features, 5 categorical features, and 19 dense features. We use the last interacted list as the target, while interactions from previous lists are used to build the history lists.
 \item \textbf{ Avito\footnote{\url{https://www.kaggle.com/c/avito-context-ad-clicks/data}}}: A dataset consists of multiple user search logs and metadata. For each user, ads on the most recent search page are used as prediction targets, while the preceding pages serve as historical behaviors.

\end{itemize}
In terms of data processing, we adopted a similar approach to previous studies~\cite{MIR,liu2022neural,huang2024chemeval,shenp,huang2025selfaug,liang2025adaptive,pan2025revisiting}, removing users and items with fewer than four interaction records and retaining only users with at least one click, ensuring sufficient interaction data to accurately capture user preferences. The detailed statistical information of the processed datasets is summarized in Table \ref{tab:dataset}.

\subsubsection{Evaluation Metrics}
Reranking encompasses more diverse objectives, such as utility~\cite{DLCM}, diversity~\cite{carraro2024enhancing}, and fairness~\cite{han2023fair}. This study focuses on utility-driven reranking, and we adopt the widely used NDCG@K and MAP@K~\cite{MIR,jarvelin2002cumulated,yin2024dataset,ye2025fuxi,yin2025feature,yu2025thought,ye2025fuxi2}. Since the number of reranked items in real-world applications is typically in the tens~\cite{relife,PRM}, we set the length of the reranked result list to 10 for the Ad and PRM Public datasets, and accordingly set $K = 5$ and $10$. Since the candidate set in the Avito dataset contains only 5 items, we set $K = 5$ for this dataset~\cite{nar4rec}.

\subsubsection{Baseline Methods}
We compare the proposed method with 6 state-of-the-art reranking methods. We select miDNN, DLCM, and PRM as typical baselines representing different structures, and Seq2Slate, CRUM, PIER, and NAR4Rec as typical baselines for generative models. The details are listed as follows:
\begin{itemize}[leftmargin=*,align=left]
    \item \textbf{MIDNN}~\cite{MIDNN}:
    MIDNN extracts mutual influence between items in the input ranking list with global feature extension.
    \item \textbf{DLCM}~\cite{DLCM}:
    DLCM first employs GRU to sequentially encode the entire ranking list, generating item representations.
    \item \textbf{PRM}~\cite{PRM}:
    PRM employs self-attention to model the mutual influence between any pair of items and users’ preferences.
    \item \textbf{Seq2Slate}~\cite{seq2slate}:
    Seq2Slate is based on an RNN pointer network architecture that sequentially predicts the next item to generate the reranked list.
    \item \textbf{CRUM}~\cite{crum}:
    CRUM achieves high utility in pairwise reranking by using a utility evaluator to guide item pair swapping.
    \item \textbf{PIER}~\cite{PIER}:
    PIER utilizes a hashing algorithm to select the top-k candidates from the full permutation based on user interests. Subsequently, the generator and evaluator are jointly trained to generate more optimal permutations.
    \item \textbf{NAR4Rec}~\cite{nar4rec}: NAR4Rec employs a non-autoregressive generative model to mitigate inference bias and enhance efficiency.
\end{itemize}

\subsubsection{Hyper-parameter Settings}
We use Adam as the optimizer to implement our model and baseline methods, with a batch size of 256. We use the last 50 items clicked by the user as their history. According to different scenarios, the maximum candidate set length for Ad is set to 20 with a learning rate of 1e-4, while for PRM Public, the maximum candidate set length is set to 30 with a learning rate of 3e-5, and for Avito, the maximum candidate set length is set to 5 with a learning rate of 3e-5. The embedding size of sparse features is set to 16. Our source code will be made publicly available after the paper is published. 

\begin{table}[t]
\centering
\fontsize{8.5pt}{11pt}\selectfont
\renewcommand{\arraystretch}{1.2}
\caption{Overall performance comparison of all methods in terms of NDCG@K and MAP@K (K=5, 10). (p-value < 0.05)}
\begin{tabular}{c|cccc|cccc|cc}
    \toprule
    \small
    Datasets & \multicolumn{4}{c|}{\textbf{Ad}} & \multicolumn{4}{c|}{\textbf{PRM Public}}  & \multicolumn{2}{c}{\textbf{Avtio}} \\ 
    Metric & \textbf{N@5} & \textbf{N@10} & \textbf{M@5} & \textbf{M@10} & \textbf{N@5} & \textbf{N@10} & \textbf{M@5} & \textbf{M@10} & \textbf{N@5} & \textbf{M@5}\\
    \midrule
    miDNN & 0.6819 & 0.6952 & 0.6025 & 0.6063 & 0.2620 & 0.3341 & 0.2842 & 0.2954 & 0.6737 & 0.6838\\ 
    DLCM & 0.6857 & 0.6989 & 0.6064 & 0.6100 & 0.2664 & 0.3395 & 0.2918 & 0.3023 & 0.6912 & 0.7057\\ 
    PRM & 0.6872 & 0.7015 & 0.6080 & 0.6117 & 0.2689 & 0.3430 & 0.2937 & 0.3042 & 0.7047 & 0.7174\\ 
    Seq2Slate & 0.6916 & 0.7062 & 0.6119 & 0.6155 & 0.2709 & 0.3449 & 0.2959 & 0.3061 & 0.7245 & 0.7360\\
    CRUM & 0.6899 & 0.7037 & 0.6084 & 0.6112 & 0.2703 & 0.3445 & 0.2952 & 0.3056 & 0.7110 & 0.7266 \\
    PIER & 0.6940 & 0.7070 & 0.6123 & 0.6158 & 0.2694 & 0.3423 & 0.2941 & 0.3038 & 0.7156 & 0.7292\\
    NAR4Rec & \underline{0.6965} & \underline{0.7101} & \underline{0.6151} & \underline{0.6188} & 0.2701 & 0.3455 & 0.2968 & 0.3071 & \underline{0.7315} & \underline{0.7462}\\
    
    \textbf{PSAD-G} & \textbf{0.7038} & \textbf{0.7184} & \textbf{0.6216} & \textbf{0.6255} & \textbf{0.2762} & \textbf{0.3528} & \textbf{0.3039} & \textbf{0.3136} & \textbf{0.7427} & \textbf{0.7561}\\
    \textbf{PSAD-S} & 0.6962 & 0.7091 & 0.6146 & 0.6185 & \underline{0.2717} & \underline{0.3464} & \underline{0.2985} & \underline{0.3079} & 0.7285 & 0.7396 \\
    \bottomrule
\end{tabular}
\vspace{-2mm}
\label{tab: allperformance}
\end{table}
\subsection{Overall Performance Comparison}
For all experiments, we first train the generator and score network. Refer to Section \ref{model_inference}, for our approach, the variant that performs inference solely with the trained generator is denoted as \textbf{PSAD-G}, while the variant that jointly trains the generator and scorer and performs inference via the online distilled scoring network is denoted as \textbf{PSAD-S}.

\subsubsection{Performance Comparison}
In this part, we compare PSAD against both discriminative and generative reranking baselines, with the results presented in Table \ref{tab: allperformance}. We also conduct Wilcoxon signed rank tests~\cite{derrick2017comparing} to evaluate the statistical significance of PSAD with the base model. We have the following observations:

(1) The generative paradigm outperforms the discriminative paradigm. Even the earlier generative model Seq2Slate surpasses strong discriminative baselines such as PRM, highlighting the importance of the research problem we address. By leveraging generative methods to model the mutual influences among items within the final reranked list, we can more effectively capture contextual dependencies and achieve superior ranking performance.

(2) Methods that model global relationships among candidate items (e.g., PRM) outperform the pointwise method MIDNN and unidirectional approaches DLRM, highlighting the effectiveness of our choice to employ attention mechanisms for capturing inter-item dependencies.

(3) The generative model PSAD-G consistently outperforms other generative methods across all datasets, demonstrating strong effectiveness and generalization ability. This is attributed to the semi-autoregressive generation strategy and contextual enhancement module, which jointly improve generation quality. Moreover, the proposed User Profile Network efficiently leverages user representations, enhancing personalized modeling.

(4) The scoring model PSAD-S outperforms all discriminative baselines and performs comparably to the strongest generative baseline NAR4Rec. This confirms that our online knowledge distillation framework effectively compresses the generative model’s item distribution into a lightweight scoring network. Furthermore, the subsequent efficiency comparison validates the framework’s superior efficiency in both training and inference phases.

\begin{table}[t]
\centering
\fontsize{8.5pt}{11pt}\selectfont
\renewcommand{\arraystretch}{1.2}
\caption{The training and inference time comparison between
 PSAD and baseline methods on the PRM Public dataset. The comparison is conducted under different batch sizes, and the time is calculated by averaging the results over 10 steps. The time unit is seconds.}
 \label{tab:ablation}
 \vspace{-2mm}
\begin{tabular}{c|cc|cc|cc}
\toprule
\textbf{Batch Size} & \multicolumn{2}{c|}{\textbf{64}} & \multicolumn{2}{c|}{\textbf{128}} &  \multicolumn{2}{c}{\textbf{256}}\\ 
\textbf{Model} & \textbf{Train(↓)} & \textbf{Inf.(↓)} & \textbf{Train(↓)} & \textbf{Inf.(↓)} & \textbf{Train(↓)} & \textbf{Inf.(↓)} \\ \hline
miDNN  & 0.0895 & 0.0764 & 0.1425 & 0.1256 & 0.2132 & 0.1893  \\ 
DLCM   & 0.2799 & 0.1497 & 0.4301 & 0.2409 & 0.6827 & 0.3940  \\
PRM    & 0.1238 & 0.1014 & 0.1852 & 0.1360 & 0.2806 & 0.2037  \\
Seq2Slate  & 0.3142 & 0.2176 & 0.5114 & 0.3993 & 0.8384 & 0.5873  \\
CRUM   & 0.2021 & 0.1589 & 0.3185 & 0.2850 & 0.4415 & 0.3048  \\
PIER   & 0.2428 & 0.2247 & 0.3608 & 0.3394 & 0.5482 & 0.4735  \\
NAR4Rec  & 0.1903 & 0.1356 & 0.2904 & 0.2215 & 0.4649 & 0.3572  \\ 
\textbf{PSAD-G}  & 0.1834 & 0.1568 & 0.2860 & 0.2376 & 0.4931  & 0.4239  \\
\textbf{PSAD-S}  & 0.1897 & 0.0801 & 0.2943 & 0.1290 & 0.5128  & 0.2161  \\
\bottomrule
\end{tabular}
\vspace{-3.5mm}
\end{table}
\subsubsection{Efficiency Comparison}
Following prior work~\cite{nar4rec,guo2024scaling,gu2025rapid,xu2025multi,xie2024breaking,xie2025breaking}, we measure the average training and inference time and compare our model with baselines under varying batch sizes on the PRM Public dataset, as shown in Table \ref{tab:ablation}:

(1) Generative models exhibit significantly higher training and inference time consumption compared to discriminative models, which validates our research motivation that existing generative approaches often struggle to balance performance and efficiency, leading to substantial computational overhead.

(2) PSAD demonstrates markedly higher training efficiency than the autoregressive generative model Seq2Slate, while remaining comparable to non-autoregressive models such as NAR4Rec and CRUM. This confirms that the semi-autoregressive generation strategy adopted by PSAD effectively maintains high-quality generation without introducing excessive training costs.

(3) PSAD-S achieves faster inference than all generative baselines and even surpasses some complex discriminative models such as PRM. Meanwhile, its overall training cost shows only a marginal increase compared to PSAD-G. These results validate the effectiveness of our shared-encoder online distillation framework, which introduces a lightweight scoring model with minimal additional training cost while achieving extremely low inference latency.

\begin{table}[t]
\centering
\fontsize{8.5pt}{11pt}\selectfont
\renewcommand{\arraystretch}{1.2}
\caption{Ablation analysis of PSAD-G}
\label{tab:abs}
\vspace{-2mm}
\begin{tabular}{c|cc|cc|cc}
\toprule
Datasets & \multicolumn{2}{c|}{\textbf{Ad}} & \multicolumn{2}{c|}{\textbf{PRM Public}} &  \multicolumn{2}{c}{\textbf{Avito}}\\ 
Metric &  \textbf{N@10}  & \textbf{M@10}  & \textbf{N@10} & \textbf{M@10} & \textbf{N@5} & \textbf{M@5}\\ \hline
\textbf{\textit{w/o sa}}  & 0.7095 & 0.6192 & 0.3455 & 0.3054 & 0.7145 & 0.7295\\  
\textbf{\textit{w/o ce}}  & 0.7133 & 0.6220 & 0.3482 & 0.3096 & 0.7231 & 0.7360\\
\textbf{\textit{w/o sace}}  & 0.7042 & 0.6139 & 0.3441 & 0.3052 & 0.7103 & 0.7217\\
\textbf{\textit{w/o ppe}} & 0.7168 & 0.6248 & 0.3523 & 0.3128 & 0.7384 & 0.7546\\
\textbf{\textit{w/o pg}} & 0.7136 & 0.6224 & 0.3507 & 0.3110 & 0.7402 & 0.7543\\ 
\textbf{PSAD-G} & \textbf{0.7184} & \textbf{0.6255} & \textbf{0.3528} & \textbf{0.3136} & \textbf{0.7427} & \textbf{0.7561} \\ 
\bottomrule
\end{tabular}
\end{table}
\subsection{Ablation study}
We designed multiple variants of the model to investigate the effectiveness of each component in our approach and conducted a series of experiments on public datasets. 
\textbf{\textit{Variant without Semi-Autoregressive (w/o sa)}} indicates that we generate all items in one step instead of using a block-wise generation strategy, \textbf{\textit{Variant without Contextual Enhancement (w/o ce)}} means that we remove the contextual enhancement process\textbf{\textit{Variant without sa and ce (w/o sace)}} indicates that both the semi-autoregressive generation and the contextual enhancement module are removed
,  \textbf{\textit{Variant without Personalized Position Encoding (w/o ppe)}} denotes that we remove the personalized relative positional encoding used to model interest decay, \textbf{\textit{Variant without Personalized Gate (w/o pg)}} indicates that we remove the personalized gating network. 

The comparison of the above variants and the PSAD on the three public datasets is shown in Table \ref{tab:abs}:
The (\textbf{\textit{w/o sa}}) shows that compared to one-shot generation, block-wise generation better preserves dependencies within the generated sequence, validating the effectiveness of our proposed semi-autoregressive generation.
The (\textbf{\textit{w/o ce}}) demonstrates that contextual enhancement helps enhance local coherence and inter-block consistency in the generated sequence.
The (\textbf{\textit{w/o sace}}) indicates that the combination of semi-autoregressive generation and contextual enhancement yields better performance.
The (\textbf{\textit{w/o ppe}}) reveals the importance of Personalized Position Encoding in modeling personalized interest decay, which contributes to more accurate user intent positioning in the ranking list.
The (\textbf{\textit{w/o pg}}) confirms that Personalized Gate improves user-item alignment by dynamically adjusting item features based on user profiles.


\subsection{Hyper-parameter Study}

\begin{figure}[t]
            \centering
            \setlength{\abovecaptionskip}{-5pt}   
            \setlength{\belowcaptionskip}{0pt}   
            \includegraphics[width=\textwidth]{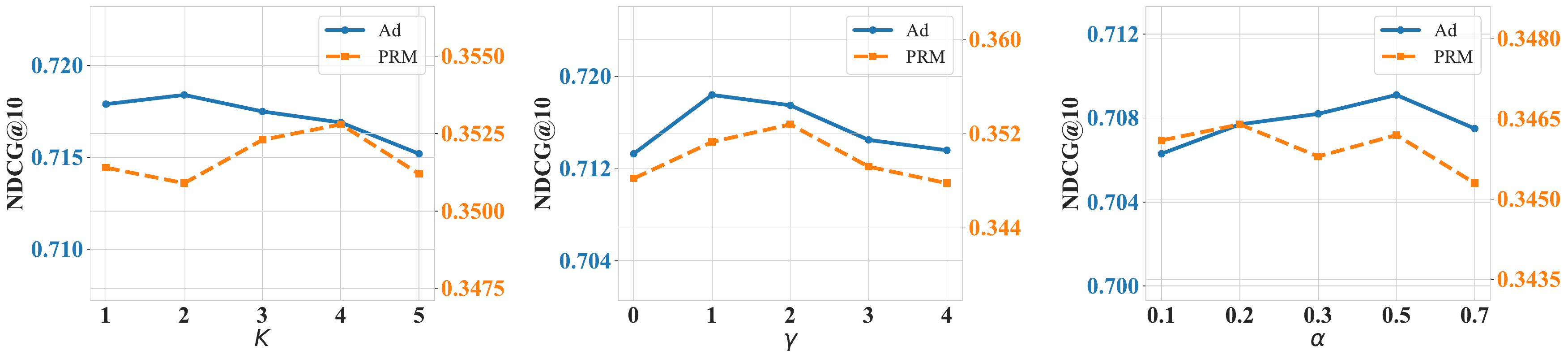}
            \vspace{-2mm}
            \caption{Hyper-Parameter Performance}
        \label{fig:hyperstudy}
        \vspace{-2mm}
\end{figure}


As shown in Fig.~\ref{fig:hyperstudy}, we analyze the key hyperparameters of PASD on two datasets, including the parallel generation block size ($K$), the number of random masks ($\gamma$), and the distillation weight ($\alpha$).
For the block size $K$, the overall performance exhibits a rise-then-fall trend. A small $K$ fails to alleviate the error accumulation from item-by-item generation, whereas an excessively large $K$ harms the internal consistency of the sequence due to overly strong independence assumptions.
For the number of random masks $\gamma$, the model achieves optimal performance when $\gamma = 1$ or $2$, while larger values disrupt the existing reasonable sequence structure.
For the distillation weight $\alpha$, PSAD-S performs best at $\alpha = 0.5$ and $0.2$ on the respective datasets. As $\alpha$ controls the trade-off between supervision from ground-truth labels and distributional guidance from the generator, suboptimal values disrupts this balance and leads to worse results.


\subsection{In-depth Analysis}
\begin{table}[t]
\centering
\fontsize{8.5pt}{11pt}\selectfont
\renewcommand{\arraystretch}{1.2}
\caption{Comparison of different distillation methods in terms of performance (NDCG@10) and training time. The batch size is set to 256, and the time is reported as the average over 10 training steps. The time unit is seconds.}
 \label{tab:dlc}
 \vspace{-2mm}
\begin{tabular}{c|cc|cc}
\toprule
\textbf{Dataset} & \multicolumn{2}{c|}{\textbf{Ad}} & \multicolumn{2}{c}{\textbf{PRM}} \\ 
\textbf{Teacher Model} & \textbf{N@10(↑)} & \textbf{Train(↓)}  & \textbf{N@10(↑)} & \textbf{Train(↓)} \\ \hline

\textbf{AR(online)} & 0.7078 & 0.4896 & 0.3434 & 0.7372\\ 

\textbf{NAR(online)} & 0.7046 & 0.3004 & 0.3429 &  0.4165\\

\textbf{SAR(offline)} &0.7083 & 0.6547 & 0.3481 & 0.8933 \\

\textbf{SAR(online)} & 0.7091 & 0.3752 & 0.3464 & 0.5128 \\

\bottomrule
\end{tabular}
\vspace{-3.5mm}
\end{table}
\begin{figure}[t]
            \centering
            \setlength{\abovecaptionskip}{-5pt}   
            \setlength{\belowcaptionskip}{0pt}   
            \includegraphics[width=0.7\textwidth]{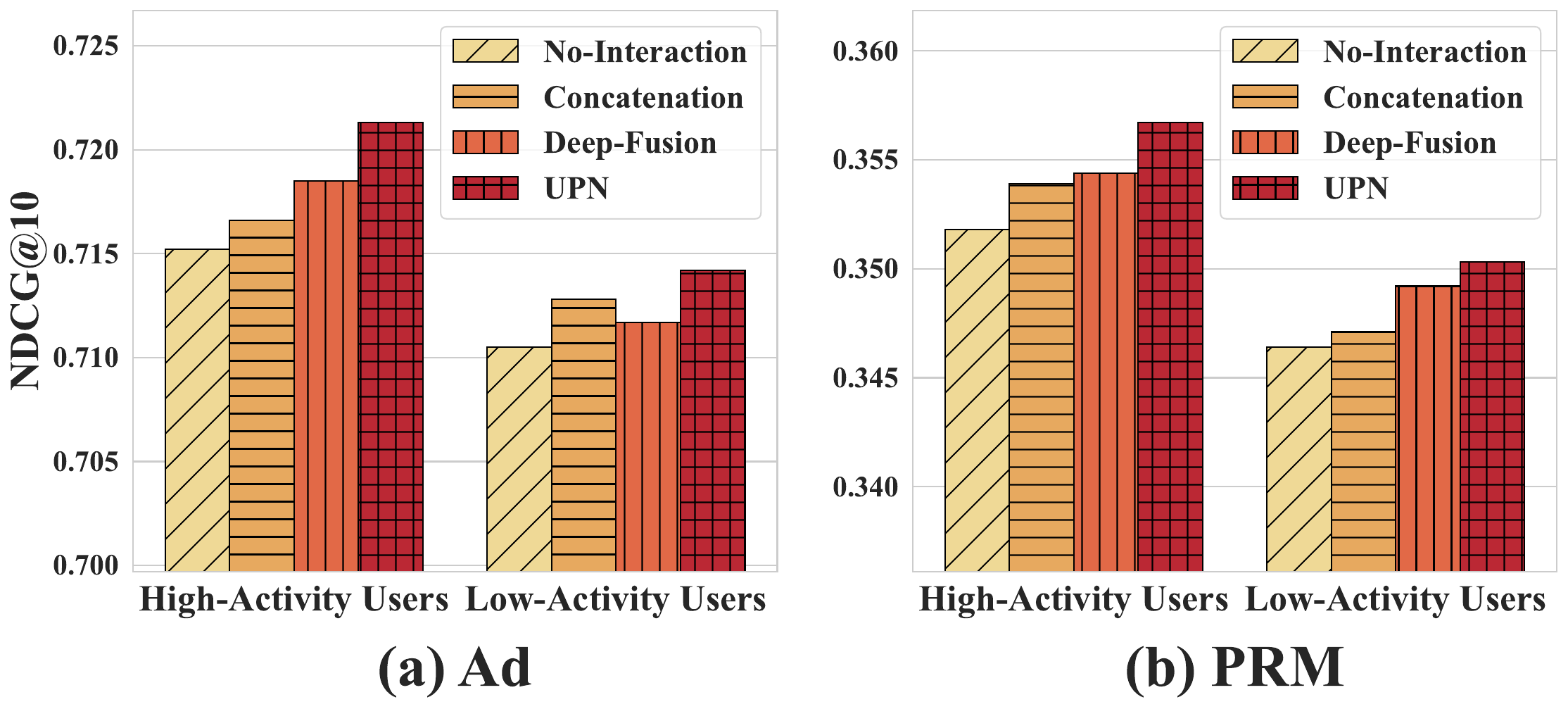}
            
            \caption{Performance comparison of item–user feature interaction mechanisms across user sequences with varying activity levels.}
        \label{fig: vis}
\end{figure}
\subsubsection{Analysis of Distillation Methods} To validate the necessity of employing a semi-autoregressive generator for online distillation of the score network, we compare the effects of different teacher models, including autoregressive(AR), non-autoregressive(NAR), and semi-autoregressive(SAR), as well as offline distillation, on the final performance of the student model. According to the results in Table~\ref{tab:dlc}, we draw the following conclusions: (1) Using a semi-autoregressive model as the teacher achieves the best distillation results while requiring significantly less training time compared with a fully autoregressive model, confirming the rationale for adopting a semi-autoregressive generator as the teacher. (2) The traditional two-stage offline distillation takes much longer than our online approach. More importantly, online distillation enables the student to learn from a continuously improving teacher rather than static knowledge, leading to comparable or even superior performance of the student model.


\subsubsection{Analysis of User–Item Interactions
}To evaluate the personalized modeling capability of the UPN module, we define the top 20\% of users based on interaction sequence length as high-activity users and the remaining users as low-activity users. We compare the performance of four interaction strategies: no interaction, concatenation~\cite{PRM}, late fusion~\cite{MIR}, and UPN, across two datasets. As shown in Fig.~\ref{fig: vis}, results indicate that UPN consistently outperforms other interaction strategies, achieving particularly pronounced improvements for high-activity users. 
This demonstrates that when sufficient personalized signals are available, the dynamic and multidimensional interaction mechanism of UPN can more effectively capture user characteristics and translate them into notable improvements in ranking performance. 



\section{Conclusion}
In this study, we investigated two major challenges in existing generative reranking methods: the trade-off between high generation quality and low inference latency, and the insufficient interaction between user and item features. To address these issues, we proposed a novel reranking framework, Personalized Semi-Autoregressive with Online Knowledge Distillation (PSAD ), which unified quality, efficiency, and personalization.
Specifically, a powerful semi-autoregressive teacher model, PSAD-G, ensured high-quality and deeply personalized ranking knowledge through a block-wise generation mechanism and the proposed User Profile Network (UPN). Meanwhile, a lightweight student network, PSAD-S, learned this knowledge in real time via online knowledge distillation, achieving fast and efficient inference.
Extensive experiments conducted on three large-scale datasets demonstrated that PSAD-G established a new benchmark in ranking performance, while PSAD-S significantly reduced inference latency with comparable accuracy. These results verified the effectiveness of PSAD in bridging the gap between generation quality and inference efficiency, while enhancing personalized reranking performance. In future work, we will further explore and optimize the distillation mechanism to enhance knowledge transfer and generalization between the two modeling paradigms.

\bibliographystyle{splncs04}
\bibliography{Sections/main}

\end{document}